\documentclass[nonacm,screen]{acmart}
\AtBeginDocument{%
  \providecommand\BibTeX{{%
    \normalfont B\kern-0.5em{\scshape i\kern-0.25em b}\kern-0.8em\TeX}}}

\setcopyright{acmcopyright}
\copyrightyear{2022}
\acmYear{2022}

\acmConference[FAccT '22]{FAccT '22: ACM Conference on Fairness, Accountability, and Transparency}{June 21--24, 2022}{Seoul, South Korea}
\acmBooktitle{FAccT '22: ACM Conference on Fairness, Accountability, and Transparency, June 21--24, 2022, Seoul, South Korea}

\usepackage[section]{placeins}

\begin{document}

\title{What Type of Explanation Do Rejected Job Applicants Want? Implications for Explainable AI}

\author{Matthew Olckers}
\email{m.olckers@unsw.edu.au}
\orcid{0001-7096-7047}
\affiliation{%
  \institution{UNSW Sydney}
  \streetaddress{P.O. Box 1212}
  \city{Sydney}
  \state{New South Wales}
  \country{Australia}
  \postcode{2052}
}
\author{Alicia Vidler}
\email{a.vidler@unsw.edu.au}
\affiliation{%
  \institution{UNSW Sydney}
  \streetaddress{P.O. Box 1212}
  \city{Sydney}
  \state{New South Wales}
  \country{Australia}
  \postcode{2052}
}
\author{Toby Walsh}
\email{tw@cse.unsw.edu.au}
\affiliation{%
  \institution{UNSW Sydney}
  \streetaddress{P.O. Box 1212}
  \city{Sydney}
  \state{New South Wales}
  \country{Australia}
  \postcode{2052}
}

\renewcommand{\shortauthors}{Olckers, Vidler, and Walsh}

\begin{abstract}
    Rejected job applicants seldom receive explanations from employers. Techniques from Explainable AI (XAI) could provide explanations at scale. Although XAI researchers have developed many different types of explanations, we know little about the type of explanations job applicants want. We use a survey of recent job applicants to fill this gap. Our survey generates three main insights. First, the current norm of, at most, generic feedback frustrates applicants. Second, applicants feel the employer has an obligation to provide an explanation. Third, job applicants want to know why they were unsuccessful and how to improve.
\end{abstract}



\begin{teaserfigure}
  \includegraphics[width=\textwidth]{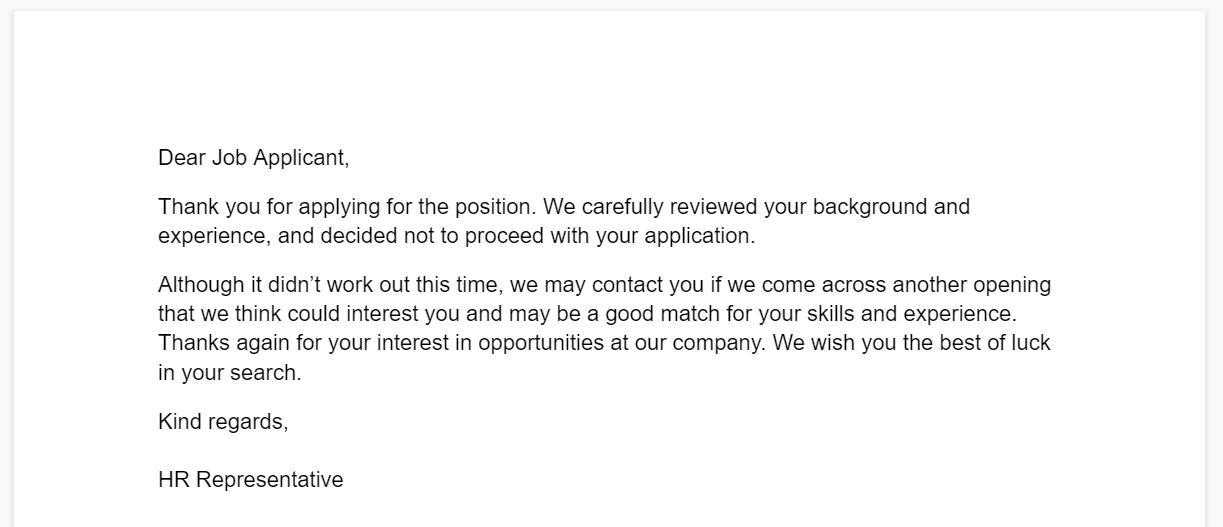}
  \label{fig:teaser}
\end{teaserfigure}

\maketitle

\section{Introduction}

Job applications can be painful, especially when your application is rejected. In return for many hours spent formatting your CV, rewriting your cover letter, and chasing up references, all too often, you simply receive a message telling you that you were unsuccessful---if you receive any message at all. You want to know why. Why did they reject your application? Who was a better fit? What should you have done differently? Unfortunately, employers do not have the time to explain their decision to every applicant.

New tools from Explainable AI (XAI) could solve this problem. An algorithm can generate explanations in seconds that would take a hiring manager hours. Although tools from XAI can generate many different types of explanations, we have little idea of what type of explanations people want.

In this paper, we use a survey of recent job applicants to discover what types of explanations applicants want. The survey provides the following insights:
\begin{itemize}
    \item The current norm of, at most, generic feedback frustrates applicants and 4 in 10 applicants state they would not apply to the same employer again if they did not receive an explanation. 
    \item In exchange for the effort of preparing a job application, applicants expect the employer to provide a personalized explanation.
    \item Job applicants want personalized feedback that explains what they should have done differently to get the job. They are also interested in the characteristics of the successful applicants and the main factors the employer used to distinguish between applicants.
\end{itemize}

Researchers have developed many techniques to explain AI decisions \citep{Adadi2018, Guidotti2018, Verma2020}, and yet the existing literature provides little information about the kinds of explanations end-users want. We can draw insights from social science research on how people present explanations to each other \citep{Miller2019, Mittelstadt2019, Srinivasan2020}, but these general insights may not provide a complete description of the ideal explanation in a specific context. Rather than taking a general approach, we study the specific context of rejected job applications.  

We study this specific context because providing explanations to rejected job applicants is an ideal application for XAI. Since a single position often receives hundreds of applicants, hiring managers do not have the time to provide personalized explanations to each applicant. Techniques from XAI allow for personalized explanations at scale. Computer systems and AI are used in many parts of the job hiring process \citep{Albert2019} so adding automatically generated explanations should be straightforward. If employers use AI to make an automated hiring decision, such as screening CVs, applicants may have a legal right to an explanation.\footnote{The European Union's General Data Protection Regulation makes provision for a right to an explanation \citep{Selbst2017}.}

Despite the potential for XAI to provide explanations to rejected job applicants, existing research provides little guidance on the types of explanations applicants want.\footnote{For example, a recent field experiment studies the timing of and the salutations used in rejection letters but does not delve into the content \citep{Cortini2019}.} The most relevant insights stem from studies in organizational and personnel psychology. Most of these studies focus on the applicants' perceptions of fairness. For example, providing information about the successful applicant and sharing when the position was withdrawn improved applicants' perception of fairness \citep{Gilliland2001}. Providing test scores relative to the successful applicants also improved perceptions of fairness but damaged rejected applicants' perceptions of themselves \citep{Ployhart1999, Schinkel2004}. These studies use tweaks in rejection letters to study the impact on applicants' reactions. Rather than using an experiment, we survey recent job applicants to discover what they would like a rejection letter to contain.

To our knowledge, our survey respondents' desire for personalized explanations has not been documented in previous studies. The survey responses show that applicants want to know why they were unsuccessful and what they can do to improve. Applicants also want to know about the characteristics of the successful applicants and the main factors employers use to make decisions, but this information received less emphasis in the open-ended responses.

We also document that applicants may be discouraged from applying for another role at the same employer if the employer does not explain a rejection decision. Earlier research has documented similar behavior \citep{Truxillo2009}. Rapid feedback increases the likelihood of future applications \citep{Cortini2019} and female applicants are less likely to apply again after being rejected \citep{Brands2017}.

Our most surprising finding is that despite the respondent's strong desire to receive explanations, 44 percent of respondents would not be willing to pay to receive an explanation. Open-ended responses show that respondents feel the employer has an obligation to provide an explanation.\footnote{A similar result was found with job applicants in the United States \citep{Waung2007}. Applicants who failed to receive a rejection notification were more likely to report that the employer has failed to meet their obligation.} Paying for an explanation to a rejected job application may be an example of a repugnant transaction \citep{Roth2007, Leuker2021}. Just the idea of attaching a monetary value to an explanation causes disgust.

In summary, our paper makes three main contributions. First, given that job rejections is a natural application of XAI, we ask job applicants directly about what they would like an explanation to contain. The survey respondents placed a strong emphasis on personalized explanations to explain why they were unsuccessful and how to improve. Second, we document a strong desire to receive explanations and a belief that employers have an obligation to provide explanations. Third, we provide evidence that paying for an explanation is a repugnant transaction. 

\section{Survey}

\subsection{Sample}

We partnered with FlexCareers, an established Australian online job platform, to distribute the survey. As the name suggests, FlexCareers was founded to champion flexible job opportunities. The platform connects over 180 thousand job candidates to hundreds of employers. Besides job matching, FlexCareers consults employers on flexible work practices. 

Of the 40 thousand monthly visitors to the FlexCareers platform, 68 percent are female and 62 percent are between 25 and 45 years of age. Of FlexCareers' Facebook followers, 93 percent are female, and 86 percent have a tertiary qualification.\footnote{Statistics reported by FlexCareers in December 2021.} Employers who use the FlexCareers platform include large Australian banking, consulting, engineering, and utility companies. 

FlexCareers emailed an invitation to complete our survey to their members who had applied for a job through the FlexCareers platform in the past year. Of the nearly 13 thousand people who were sent the email, 141 completed our survey. The response rate may be low because we did not provide any reward to complete the survey.

To ensure the survey responses were anonymous, the only demographic information we collected was the employment status and years of work experience for each respondent. Of the 141 respondents, 40 percent indicated they were unemployed. The high number of unemployed people in our sample likely stems from our decision to only email FlexCareers members who had applied for a job through the FlexCareers platform in the past year. Close to half of the sample has more than 20 years of work experience.

The lack of detailed demographic information prevents us from determining if our sample is representative of the population of Australian job seekers. We preferred to use a short and anonymous survey to take the first step in understanding what kind of explanations job applicants want. The similarity of responses suggests we have discovered some key themes that may hold more generally. However, we caution researchers from extrapolating the results to different contexts.

\subsection{Questions}

Our anonymous survey contained a mix of multiple choice and free text questions. See the appendix for the full list of survey questions. The median respondent took approximately 5 minutes to complete the survey. The Qualtrics survey design and the data are available on \href{https://researchbox.org/}{Research Box}, a secure data repository that allows for easy replication of our analysis. 


\section{Insights}

\subsection{Frustration with current practices}

Our first insight is that the current practice of either no feedback or a generic response causes frustration and disappointment for job applicants. These strong negative emotions can impact employers by reducing the future applicant pool and damaging perceptions of the employer's brand.

\subsubsection*{Lack of response causes frustration and disappointment}

Employers rarely provide feedback to rejected applicants and the feedback that is provided is often generic. Only 29 percent of the survey respondents said the employer explained why their application was unsuccessful.

For the remaining 71 percent of respondents who did not receive any explanation, we asked ``How did you feel when you did not receive an explanation for an unsuccessful job application?". Most respondents expressed strong negative feelings, particularly frustration and disappointment. Only 14 of the 74 responses can be categorized as neutral or indifferent.\footnote{For example, one respondent answered: \textit{``Business as usual, I did not expect to receive an explanation other than `Other applicants were more suitable..' or equivalent"}.} Of the 60 respondents who expressed negative feelings, 13 used the word "frustrated" or "frustrating", and 9 used the word "disappointed" or "disappointing". Examples of the responses expressing negative feelings are shown below.
\begin{itemize}
\addtolength\itemsep{2mm}
    \item[] \textit{``Inadequate and confused, many of the positions I was very much experienced enough for, even after an interview there was no explanation other than a stock standard rejection email."} 
    \item[] \textit{``Discarded. Treated like number."} 
    \item[] \textit{``Frustrated that generic answers are given even if you do get a response, completely useless, no tailored communication"} 
    \item[] \textit{``Crushed. Doubted my competence and value"}  
    \item[] \textit{``Frustrated. It happens so often that you feel like employers don't even care, you're just a number."} 
    \item[] \textit{``Extremely sad."} 
    \item[] \textit{``Depressed, unsure of reasons, uncertainty with quality of CV and skills or experience."} 
    \item[] \textit{``Disappointed and useless."} 
    \item[] \textit{``I usually feel dumbfounded whenever I do not get an explanation despite emailing the interviewers after getting a job rejection."} 
\end{itemize}

\noindent Similar to the frustration of receiving no explanation, the following response highlights the frustration from receiving generic explanations. 
\begin{quote}\textit{
``One of the worst things is an answer where you can tell that no thought has been given. It is disrespectful and indicates that the person responding either cares little about your opinion or thinks that you are somehow inferior to them. Instead of telling me that other candidates are a better fit, tell me why they are a better fit and let me know what I could have done better to meet the brief."
}\end{quote}
Generic explanations are not a new phenomenon. Rejection letters written nearly 40 years ago provided the same types of vague explanations that frustrate job applicants today \citep{Jablin1984}. 

\subsubsection*{Impact of negative feelings on employers}

If employers ignore rejected applicants, most of the rejected applicants feel frustration, disappointment, and other negative emotions. How do these negative feelings impact employers? The survey responses highlight two channels. First, rejected applicants are discouraged from applying for other positions at the same employer. Second, the negative feelings damage perceptions of the employer's brand. 

We asked respondents ``If you did not receive an explanation from an unsuccessful job application, would you be willing to apply for another position at the same employer?". Just over 40 percent of respondents said no. Even though the current norm is to receive little or no explanation after a rejected application, employers who follow this norm may shrink their pool of future job applicants.

Ignoring rejected applicants can impact the employer's brand.
In response to our question ``How did you feel when you did not receive an explanation for an unsuccessful job application?", one respondent shared:
\begin{quote}\textit{
``Frustrated. Did not reflect well on the brand of that company."
}\end{quote}
At the end of our survey, we asked for final thoughts from the respondents. One of the respondents added:
\begin{quote}\textit{
``My perception of the organizations' brands was significantly impacted by the impersonal template rejection emails from 'do not reply' email addresses."
}\end{quote}
To the extent that rejected applicants also serve as customers for the employers, the lack of explanation may push these customers to competitors.

\subsection{Applicants want personalized explanations}

The frustration with current practices raises the question: what do rejected job applicants want? Our survey responses indicate job applicants want personalized explanations. Job applicants want to know what they should have done differently to get the job and how to improve their profile.

We first asked our respondents the following open-ended question: ``Suppose an employer agreed to explain why your job application was unsuccessful. What information would you like the explanation to contain?".  Most job applicants want to know why they were unsuccessful and how they can improve. Of the 130 respondents who answered the question, 53 percent wanted to know why they were unsuccessful and 27 percent asked for information on how to improve. Only 12 percent wanted to know information about other applicants---of which several wanted to know if the employer hired an internal candidate.

For the respondents who wanted to know how they could improve, some asked for isolated feedback on the interview or the presentation of their CV while others asked for more general information on skills and qualifications. Here are examples of these two types of responses:
\begin{itemize}
\addtolength\itemsep{2mm}
\item[] \textit{``Was the resume good enough for them to look at the application? Cover letter - did they look at it? If so, where did they think that it needed improvement? Were the skills as per what they were looking for? What set out applications that were successful to mine? If given a chance, would they reconsider my application if I made changes and addressed the above points?"} 
\item[] \textit{``How to improve in the future, qualifications and training needed."} 
\end{itemize}

One surprisingly common response was a demand for ``the truth". In addition to five responses that specifically asked for ``the truth", a further eight responses asked for honesty or the ``real reason". Below are the five responses that mentioned the word ``truth":
\begin{itemize}
\addtolength\itemsep{2mm}
\item[] \textit{``Truth."} 
\item[] \textit{``The truth."} 
\item[] \textit{``Truthful explanation, and not using cliche terms."} 
\item[] \textit{``The truth. No need for politically correct well-designed marketing stuff. Let's say if the reason was the too high salary expectation, then say that. Similar with the age: don't answer: overqualified."} 
\item[] \textit{``The truth. Why you didn't get the role, what set me apart from the others, where I could improve, if at all."} 
\end{itemize}

The demand for the truth suggests that rejected job applicants feel they have information hidden from them. One respondent suggested that ``due to anti-discrimination laws [they] would expect generic excuses". Perhaps, a more plausible reason is that it is easier for the employer to send out generic information that does not reveal the true reason for the rejection. The true reason may be unique to each applicant or the employer may prefer to keep the true reason secret, such as the position was withdrawn or was earmarked for an internal candidate.

After asking the respondents an open-ended question on the information they would like an explanation to contain, we asked respondents to rate the importance of different types of content, listed below. We randomized the order in which each respondent was asked to rate each type of content. 
\begin{itemize}
    \item Characteristics of the successful applicants (such as education level, years of experience, and main skills)
    \item How your profile would need to change to improve your chances.
    \item The main factors the employer used to distinguish between applicants.
    \item Gender of the successful applicant.
    \item Confirmation that the position was filled.
    \item Your ranking and the total number of applicants.
    \item Number of applicants who continued to the next round.
\end{itemize}

\begin{figure}[!htb]
    \centering
    \includegraphics[width=\textwidth]{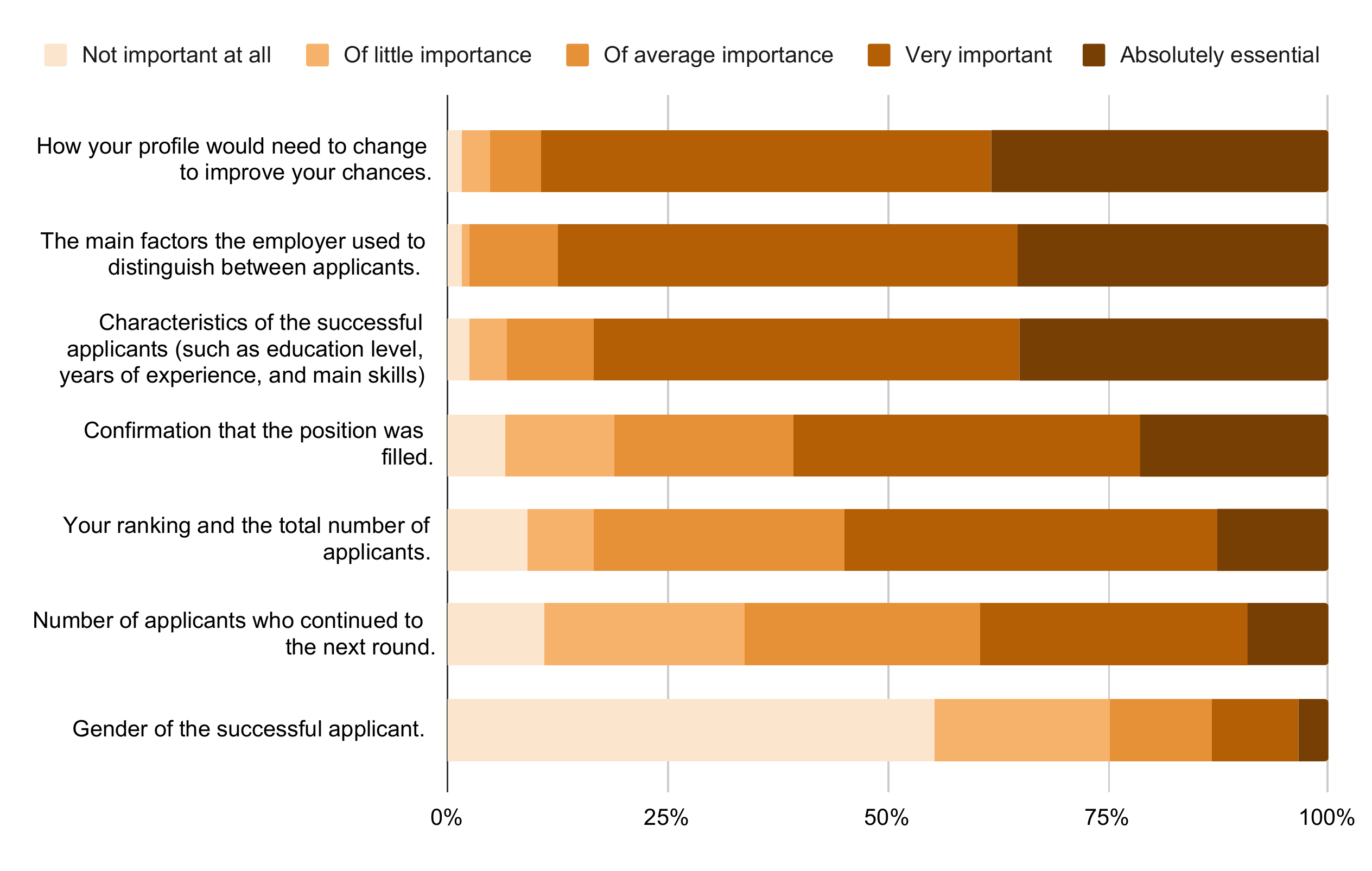}
    \caption{Respondents rated the importance of different types of content an explanation could contain}
    \label{fig:content_importance}
\end{figure}

Figure \ref{fig:content_importance} shows the ratings. Respondents gave strong weight to three types of content: how to improve, the main factors the employer used to distinguish between applicants, and characteristics of successful applicants. Each of these types of content can be implemented by techniques from XAI. Counterfactual explanations can guide job applicants on how to improve, feature importance can identify the main factors the employer used to distinguish between applications, and summary statistics can describe the characteristics of successful applicants.

In addition, we asked each respondent to pick which content type from the list above they would prefer their explanation to contain if their explanation could only contain one type of content. Respondents' choices followed the pattern in the importance ratings. Of the 133 respondents who answered this question, 39 percent chose ``how your profile would need to change to improve your chances", 26 percent chose ``the main factors the employer used to distinguish between applicants", and 24 percent chose "characteristics of the successful applicants". None of the respondents chose the gender of the successful applicant. The remaining 11 percent of respondents picked the other content types.

\FloatBarrier

\subsection{Applicants expect explanations in exchange for their effort}

The survey responses show that rejected job applicants have a strong desire to receive explanations. An explanation may allow an applicant to improve their profile, get a job earlier, or a higher-paying position. In addition to the economic benefits, the job applicant may value the emotional benefit of receiving an explanation.

Given the economic and emotional benefits of explanations, we asked the survey respondents how much they would be willing to pay for their ideal explanation. Our questions asked: 
\begin{quote}
``In a hypothetical scenario that you could purchase your ideal explanation from a past unsuccessful job application, how much would you be willing to pay for the explanation?"
\end{quote}
Respondents filled in a monetary amount in Australian Dollars (AUD).

\begin{figure}[!htb]
    \centering
    \includegraphics[width=\textwidth]{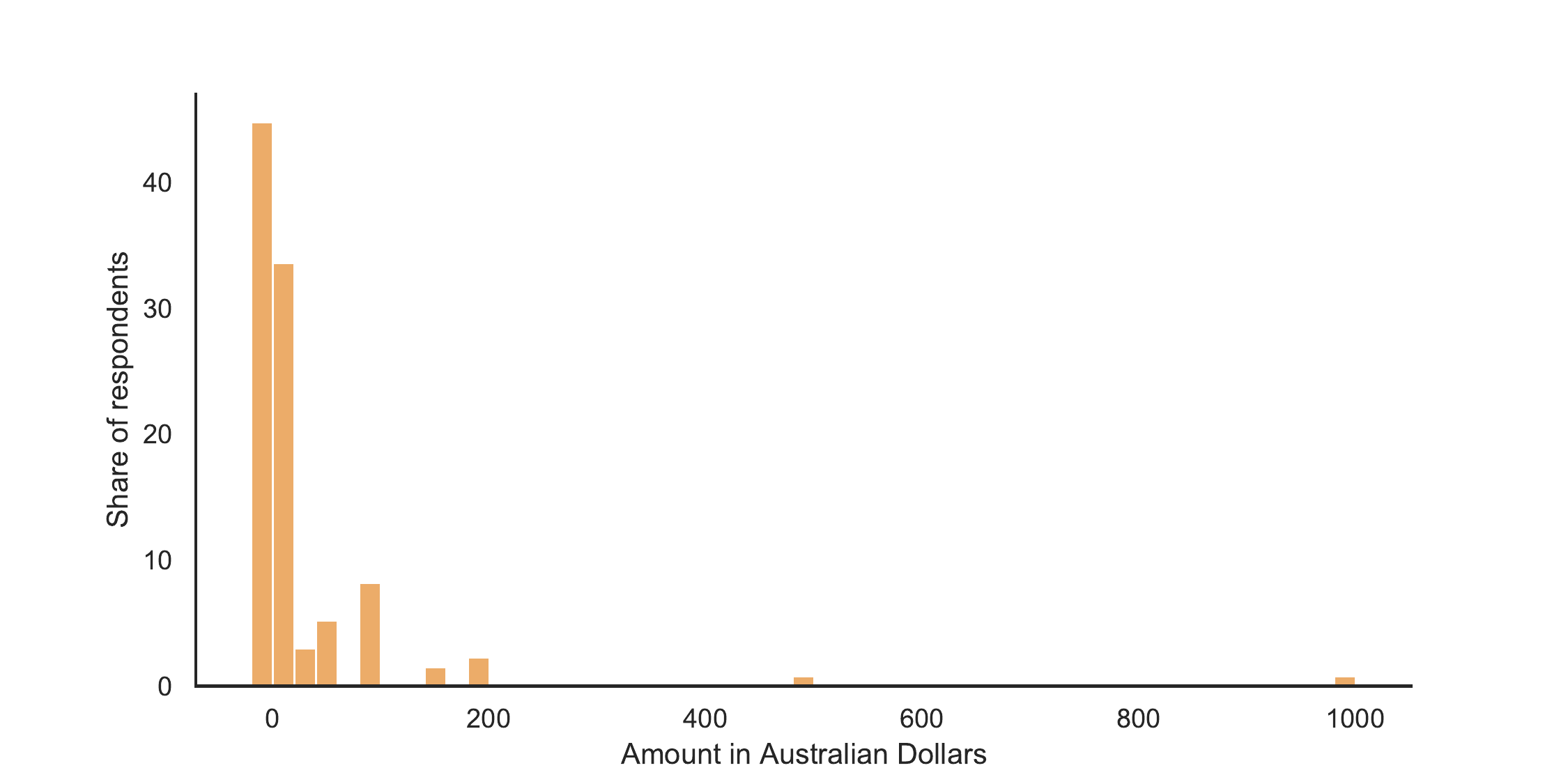}
    \caption{In a hypothetical scenario that you could purchase your ideal explanation from a past unsuccessful job application, how much would you be willing to pay for the explanation?}
    \label{fig:wtp}
\end{figure}

Figure \ref{fig:wtp} shows the amounts chosen by the respondents. Surprisingly, 44 percent of respondents selected zero. Only 25 percent of respondents were willing to pay more than 20 AUD. Our respondents showed a strong desire to receive explanations and yet these responses suggest they do not attach a high value to the explanations. A deeper look into the open-ended responses can explain this surprising result.

The open-ended responses suggest the survey respondents believe they are entitled to an explanation in exchange for their effort on the job application. They feel they deserve the explanation without paying an additional cost. The following responses provide examples of this viewpoint. We show the amount the respondent was willing to pay in brackets.
\begin{itemize}
\addtolength\itemsep{2mm}
    \item[] \textit{``Some employers make you bend over backward through the interview process. If you get rejected after jumping through so many hoops, the applicant deserves some feedback as to why, not just a generic rejection email."} (10 AUD) 
    \item[] \textit{``The idea about paying for feedback is idiotic and I beg you not to put it into the universe. If I take the time to apply for a job they should have the courtesy to provide feedback. Job hunting is hard enough and expensive don't add more cost to excuse inexcusable conduct."} (zero AUD) 
    \item[] \textit{``It is common courtesy to reply and let someone know where they stand in the application process, so many employers do not even respond even after the interview process. "} (zero AUD) 
    \item[] \textit{``People who don't have a job, don't have spare money to find out why they didn't get the job.  If a company has progressed a candidate to interview, there should be direct, interpersonal contact with a conversation about why the application didn't progress.  If a targeted application is requested by the employer, eg. where the application process requires a time investment from the applicant to answer specific questions responding to criteria, etc, but the application doesn't progress to interview, then the business should return information about what the key basis was for eliminating the applicant from the candidate pool.  If a quick application is submitted online, involving minimal effort other than attaching a resume and brief cover letter, a minimum response should be a standard template confirming the application would not progress."} (zero AUD) 
    \item[] \textit{``Recruiters need to respect the time it takes to submit applications and pay jobseekers the courtesy of replying"} (10 AUD) 
    \item[] \textit{``If a candidate invests time preparing and submitting an application the recruiter should be, not only morally, but also legally obligated to provide informed feedback to candidates."} (zero AUD) 
\end{itemize}

The passionate responses suggest that paying for an explanation to a rejected job application is an example of a repugnant transaction \citep{Roth2007, Leuker2021}. The respondents are disgusted by the idea of attaching a monetary value to an explanation.


\section{Concluding Remarks}

Currently, XAI benefits data analysts rather than end-users. Data analysts use automated explanations to check their machine learning models, but these explanations seldom reach end-users \citep{Bhatt2020}. To help bring the automated explanations to end-users, such as rejected job applicants, we took a small step---asking what the end-users want to know. We hope that this study can serve as a template for similar studies in other domains. We encourage researchers to adapt our survey, which is posted on \href{https://researchbox.org/}{Research Box}.

\begin{acks}
We thank the team from FlexCareers for collaborating with us on the survey. Our survey was approved by our university's ethics review board.
\end{acks}

\bibliographystyle{ACM-Reference-Format}
\bibliography{references}


\begin{thebibliography}{19}


\ifx \showCODEN    \undefined \def \showCODEN     #1{\unskip}     \fi
\ifx \showDOI      \undefined \def \showDOI       #1{#1}\fi
\ifx \showISBNx    \undefined \def \showISBNx     #1{\unskip}     \fi
\ifx \showISBNxiii \undefined \def \showISBNxiii  #1{\unskip}     \fi
\ifx \showISSN     \undefined \def \showISSN      #1{\unskip}     \fi
\ifx \showLCCN     \undefined \def \showLCCN      #1{\unskip}     \fi
\ifx \shownote     \undefined \def \shownote      #1{#1}          \fi
\ifx \showarticletitle \undefined \def \showarticletitle #1{#1}   \fi
\ifx \showURL      \undefined \def \showURL       {\relax}        \fi
\providecommand\bibfield[2]{#2}
\providecommand\bibinfo[2]{#2}
\providecommand\natexlab[1]{#1}
\providecommand\showeprint[2][]{arXiv:#2}

\bibitem[\protect\citeauthoryear{Adadi and Berrada}{Adadi and Berrada}{2018}]%
        {Adadi2018}
\bibfield{author}{\bibinfo{person}{Amina Adadi} {and} \bibinfo{person}{Mohammed
  Berrada}.} \bibinfo{year}{2018}\natexlab{}.
\newblock \showarticletitle{Peeking inside the black-box: a survey on
  explainable artificial intelligence (XAI)}.
\newblock \bibinfo{journal}{\emph{IEEE access}}  \bibinfo{volume}{6}
  (\bibinfo{year}{2018}), \bibinfo{pages}{52138--52160}.
\newblock


\bibitem[\protect\citeauthoryear{Albert}{Albert}{2019}]%
        {Albert2019}
\bibfield{author}{\bibinfo{person}{E Albert}.} \bibinfo{year}{2019}\natexlab{}.
\newblock \showarticletitle{AI in talent acquisition: a review of
  AI-applications used in recruitment and selection}.
\newblock \bibinfo{journal}{\emph{Strategic HR Review}} \bibinfo{volume}{18},
  \bibinfo{number}{5} (\bibinfo{year}{2019}), \bibinfo{pages}{215--221}.
\newblock
\urldef\tempurl%
\url{https://doi.org/10.1108/shr-04-2019-0024}
\showDOI{\tempurl}


\bibitem[\protect\citeauthoryear{Bhatt, Xiang, Sharma, Weller, Taly, Jia,
  Ghosh, Puri, Moura, and Eckersley}{Bhatt et~al\mbox{.}}{2020}]%
        {Bhatt2020}
\bibfield{author}{\bibinfo{person}{Umang Bhatt}, \bibinfo{person}{Alice Xiang},
  \bibinfo{person}{Shubham Sharma}, \bibinfo{person}{Adrian Weller},
  \bibinfo{person}{Ankur Taly}, \bibinfo{person}{Yunhan Jia},
  \bibinfo{person}{Joydeep Ghosh}, \bibinfo{person}{Ruchir Puri},
  \bibinfo{person}{Jos\'{e} M.~F. Moura}, {and} \bibinfo{person}{Peter
  Eckersley}.} \bibinfo{year}{2020}\natexlab{}.
\newblock \showarticletitle{Explainable Machine Learning in Deployment}. In
  \bibinfo{booktitle}{\emph{Proceedings of the 2020 Conference on Fairness,
  Accountability, and Transparency}} (Barcelona, Spain)
  \emph{(\bibinfo{series}{FAT* '20})}. \bibinfo{publisher}{Association for
  Computing Machinery}, \bibinfo{address}{New York, NY, USA},
  \bibinfo{pages}{648–657}.
\newblock
\showISBNx{9781450369367}
\urldef\tempurl%
\url{https://doi.org/10.1145/3351095.3375624}
\showDOI{\tempurl}


\bibitem[\protect\citeauthoryear{Brands and Fernandez-Mateo}{Brands and
  Fernandez-Mateo}{2017}]%
        {Brands2017}
\bibfield{author}{\bibinfo{person}{R Brands} {and} \bibinfo{person}{I
  Fernandez-Mateo}.} \bibinfo{year}{2017}\natexlab{}.
\newblock \showarticletitle{Leaning out: How negative recruitment experiences
  shape women’s decisions to compete for executive roles}.
\newblock \bibinfo{journal}{\emph{Administrative Science Quarterly}}
  \bibinfo{volume}{62}, \bibinfo{number}{3} (\bibinfo{year}{2017}),
  \bibinfo{pages}{405--442}.
\newblock
\urldef\tempurl%
\url{https://doi.org/10.1177/0001839216682728}
\showDOI{\tempurl}


\bibitem[\protect\citeauthoryear{Cortini, Galanti, and Barattucci}{Cortini
  et~al\mbox{.}}{2019}]%
        {Cortini2019}
\bibfield{author}{\bibinfo{person}{Michela Cortini}, \bibinfo{person}{Teresa
  Galanti}, {and} \bibinfo{person}{Massimiliano Barattucci}.}
  \bibinfo{year}{2019}\natexlab{}.
\newblock \showarticletitle{The Effect of Different Rejection Letters on
  Applicants’ Reactions}.
\newblock \bibinfo{journal}{\emph{Behavioral Sciences}} \bibinfo{volume}{9},
  \bibinfo{number}{102} (\bibinfo{year}{2019}), \bibinfo{numpages}{15}~pages.
\newblock
\urldef\tempurl%
\url{https://doi.org/10.3390/bs9100102}
\showDOI{\tempurl}


\bibitem[\protect\citeauthoryear{Gilliland, Groth, Baker, Dew, Polly, and
  Langdon}{Gilliland et~al\mbox{.}}{2001}]%
        {Gilliland2001}
\bibfield{author}{\bibinfo{person}{Stephen~W. Gilliland},
  \bibinfo{person}{Markus Groth}, \bibinfo{person}{Robert~C. Baker},
  \bibinfo{person}{Angela~E Dew}, \bibinfo{person}{Lisa~M. Polly}, {and}
  \bibinfo{person}{Jay~C. Langdon}.} \bibinfo{year}{2001}\natexlab{}.
\newblock \showarticletitle{Improving applicants' reactions to rejection
  letters: An application of fairness theory}.
\newblock \bibinfo{journal}{\emph{Personnel Psychology}} \bibinfo{volume}{54},
  \bibinfo{number}{3} (\bibinfo{year}{2001}), \bibinfo{pages}{669--703}.
\newblock
\urldef\tempurl%
\url{https://doi.org/10.1111/j.1744-6570.2001.tb00227.x}
\showDOI{\tempurl}


\bibitem[\protect\citeauthoryear{Guidotti, Monreale, Ruggieri, Turini,
  Giannotti, and Pedreschi}{Guidotti et~al\mbox{.}}{2018}]%
        {Guidotti2018}
\bibfield{author}{\bibinfo{person}{Riccardo Guidotti}, \bibinfo{person}{Anna
  Monreale}, \bibinfo{person}{Salvatore Ruggieri}, \bibinfo{person}{Franco
  Turini}, \bibinfo{person}{Fosca Giannotti}, {and} \bibinfo{person}{Dino
  Pedreschi}.} \bibinfo{year}{2018}\natexlab{}.
\newblock \showarticletitle{A Survey of Methods for Explaining Black Box
  Models}.
\newblock \bibinfo{journal}{\emph{Comput. Surveys}} \bibinfo{volume}{51},
  \bibinfo{number}{5}, Article \bibinfo{articleno}{93} (\bibinfo{year}{2018}),
  \bibinfo{numpages}{42}~pages.
\newblock
\showISSN{0360-0300}
\urldef\tempurl%
\url{https://doi.org/10.1145/3236009}
\showDOI{\tempurl}


\bibitem[\protect\citeauthoryear{Jablin and Krone}{Jablin and Krone}{1984}]%
        {Jablin1984}
\bibfield{author}{\bibinfo{person}{Fredric~M. Jablin} {and}
  \bibinfo{person}{Kathleen~J. Krone}.} \bibinfo{year}{1984}\natexlab{}.
\newblock \showarticletitle{Characteristics of Rejection Letters and Their
  Effects on Job Applicants.}
\newblock \bibinfo{journal}{\emph{Written Communication}} \bibinfo{volume}{1},
  \bibinfo{number}{4} (\bibinfo{year}{1984}), \bibinfo{pages}{387--406}.
\newblock
\urldef\tempurl%
\url{https://doi.org/10.1177/0741088384001004001}
\showDOI{\tempurl}


\bibitem[\protect\citeauthoryear{Leuker, Samartzidis, and Hertwig}{Leuker
  et~al\mbox{.}}{2021}]%
        {Leuker2021}
\bibfield{author}{\bibinfo{person}{Christina Leuker}, \bibinfo{person}{Lasare
  Samartzidis}, {and} \bibinfo{person}{Ralph Hertwig}.}
  \bibinfo{year}{2021}\natexlab{}.
\newblock \showarticletitle{What makes a market transaction morally repugnant?}
\newblock \bibinfo{journal}{\emph{Cognition}} \bibinfo{volume}{212},
  \bibinfo{number}{104644} (\bibinfo{year}{2021}),
  \bibinfo{numpages}{15}~pages.
\newblock


\bibitem[\protect\citeauthoryear{Miller}{Miller}{2019}]%
        {Miller2019}
\bibfield{author}{\bibinfo{person}{Tim Miller}.}
  \bibinfo{year}{2019}\natexlab{}.
\newblock \showarticletitle{Explanation in artificial intelligence: Insights
  from the social sciences}.
\newblock \bibinfo{journal}{\emph{Artificial Intelligence}}
  \bibinfo{volume}{267} (\bibinfo{year}{2019}), \bibinfo{pages}{1--38}.
\newblock
\showISSN{0004-3702}
\urldef\tempurl%
\url{https://doi.org/10.1016/j.artint.2018.07.007}
\showDOI{\tempurl}


\bibitem[\protect\citeauthoryear{Mittelstadt, Russell, and Wachter}{Mittelstadt
  et~al\mbox{.}}{2019}]%
        {Mittelstadt2019}
\bibfield{author}{\bibinfo{person}{Brent Mittelstadt}, \bibinfo{person}{Chris
  Russell}, {and} \bibinfo{person}{Sandra Wachter}.}
  \bibinfo{year}{2019}\natexlab{}.
\newblock \showarticletitle{Explaining Explanations in AI}. In
  \bibinfo{booktitle}{\emph{Proceedings of the Conference on Fairness,
  Accountability, and Transparency}} (Atlanta, GA, USA)
  \emph{(\bibinfo{series}{FAT* '19})}. \bibinfo{publisher}{Association for
  Computing Machinery}, \bibinfo{address}{New York, NY, USA},
  \bibinfo{pages}{279–288}.
\newblock
\showISBNx{9781450361255}
\urldef\tempurl%
\url{https://doi.org/10.1145/3287560.3287574}
\showDOI{\tempurl}


\bibitem[\protect\citeauthoryear{Ployhart, Ryan, and Bennett}{Ployhart
  et~al\mbox{.}}{1999}]%
        {Ployhart1999}
\bibfield{author}{\bibinfo{person}{Robert~E Ployhart},
  \bibinfo{person}{Ann~Marie Ryan}, {and} \bibinfo{person}{Matthew Bennett}.}
  \bibinfo{year}{1999}\natexlab{}.
\newblock \showarticletitle{Explanations for selection decisions: Applicants'
  reactions to informational and sensitivity features of explanations.}
\newblock \bibinfo{journal}{\emph{Journal of Applied Psychology}}
  \bibinfo{volume}{84}, \bibinfo{number}{1} (\bibinfo{year}{1999}),
  \bibinfo{pages}{87}.
\newblock


\bibitem[\protect\citeauthoryear{Roth}{Roth}{2007}]%
        {Roth2007}
\bibfield{author}{\bibinfo{person}{Alvin~E. Roth}.}
  \bibinfo{year}{2007}\natexlab{}.
\newblock \showarticletitle{Repugnance as a Constraint on Markets}.
\newblock \bibinfo{journal}{\emph{Journal of Economic Perspectives}}
  \bibinfo{volume}{21}, \bibinfo{number}{3} (\bibinfo{date}{September}
  \bibinfo{year}{2007}), \bibinfo{pages}{37--58}.
\newblock
\urldef\tempurl%
\url{https://doi.org/10.1257/jep.21.3.37}
\showDOI{\tempurl}


\bibitem[\protect\citeauthoryear{Schinkel, Schinkel, van Dierendonck, van
  Dierendonck, and Anderson}{Schinkel et~al\mbox{.}}{2004}]%
        {Schinkel2004}
\bibfield{author}{\bibinfo{person}{S. Schinkel}, \bibinfo{person}{Sonja
  Schinkel}, \bibinfo{person}{D. van Dierendonck}, \bibinfo{person}{Dirk van
  Dierendonck}, {and} \bibinfo{person}{Neil Anderson}.}
  \bibinfo{year}{2004}\natexlab{}.
\newblock \showarticletitle{The Impact of Selection Encounters on Applicants:
  An Experimental Study into Feedback Effects after a Negative Selection
  Decision}.
\newblock \bibinfo{journal}{\emph{International Journal of Selection and
  Assessment}} \bibinfo{volume}{12}, \bibinfo{number}{1-2}
  (\bibinfo{year}{2004}), \bibinfo{pages}{197--205}.
\newblock
\urldef\tempurl%
\url{https://doi.org/10.1111/j.0965-075x.2004.00274.x}
\showDOI{\tempurl}


\bibitem[\protect\citeauthoryear{Selbst and Powles}{Selbst and Powles}{2017}]%
        {Selbst2017}
\bibfield{author}{\bibinfo{person}{Andrew~D Selbst} {and}
  \bibinfo{person}{Julia Powles}.} \bibinfo{year}{2017}\natexlab{}.
\newblock \showarticletitle{{Meaningful information and the right to
  explanation}}.
\newblock \bibinfo{journal}{\emph{International Data Privacy Law}}
  \bibinfo{volume}{7}, \bibinfo{number}{4} (\bibinfo{date}{12}
  \bibinfo{year}{2017}), \bibinfo{pages}{233--242}.
\newblock
\showISSN{2044-3994}
\urldef\tempurl%
\url{https://doi.org/10.1093/idpl/ipx022}
\showDOI{\tempurl}
\showeprint{https://academic.oup.com/idpl/article-pdf/7/4/233/22923065/ipx022.pdf}


\bibitem[\protect\citeauthoryear{Srinivasan and Chander}{Srinivasan and
  Chander}{2020}]%
        {Srinivasan2020}
\bibfield{author}{\bibinfo{person}{Ramya Srinivasan} {and}
  \bibinfo{person}{Ajay Chander}.} \bibinfo{year}{2020}\natexlab{}.
\newblock \showarticletitle{Explanation Perspectives from the Cognitive
  Sciences---A Survey}. In \bibinfo{booktitle}{\emph{Proceedings of the
  Twenty-Ninth International Joint Conference on Artificial Intelligence,
  {IJCAI-20}}}, \bibfield{editor}{\bibinfo{person}{Christian Bessiere}} (Ed.).
  \bibinfo{publisher}{International Joint Conferences on Artificial
  Intelligence Organization}, \bibinfo{address}{Yokohama, Japan},
  \bibinfo{pages}{4812--4818}.
\newblock
\urldef\tempurl%
\url{https://doi.org/10.24963/ijcai.2020/670}
\showDOI{\tempurl}
\newblock
\shownote{Survey track.}


\bibitem[\protect\citeauthoryear{Truxillo, Bodner, Bertolino, Bauer, and
  Yonce}{Truxillo et~al\mbox{.}}{2009}]%
        {Truxillo2009}
\bibfield{author}{\bibinfo{person}{Donald~M Truxillo}, \bibinfo{person}{Todd~E
  Bodner}, \bibinfo{person}{Marilena Bertolino}, \bibinfo{person}{Talya~N
  Bauer}, {and} \bibinfo{person}{Clayton~A Yonce}.}
  \bibinfo{year}{2009}\natexlab{}.
\newblock \showarticletitle{Effects of explanations on applicant reactions: A
  meta-analytic review}.
\newblock \bibinfo{journal}{\emph{International Journal of Selection and
  Assessment}} \bibinfo{volume}{17}, \bibinfo{number}{4}
  (\bibinfo{year}{2009}), \bibinfo{pages}{346--361}.
\newblock


\bibitem[\protect\citeauthoryear{Verma, Dickerson, and Hines}{Verma
  et~al\mbox{.}}{2020}]%
        {Verma2020}
\bibfield{author}{\bibinfo{person}{Sahil Verma}, \bibinfo{person}{John
  Dickerson}, {and} \bibinfo{person}{Keegan Hines}.}
  \bibinfo{year}{2020}\natexlab{}.
\newblock \bibinfo{title}{Counterfactual explanations for machine learning: A
  review}.
\newblock
\newblock
\showeprint[arxiv]{2010.10596}


\bibitem[\protect\citeauthoryear{Waung and Brice}{Waung and Brice}{2007}]%
        {Waung2007}
\bibfield{author}{\bibinfo{person}{Marie Waung} {and}
  \bibinfo{person}{Thomas~S. Brice}.} \bibinfo{year}{2007}\natexlab{}.
\newblock \showarticletitle{The Effect of Acceptance/Rejection Status, Status
  Notification, and Organizational Obligation Fulfillment on Applicant
  Intentions}.
\newblock \bibinfo{journal}{\emph{Journal of Applied Social Psychology}}
  \bibinfo{volume}{37}, \bibinfo{number}{9} (\bibinfo{year}{2007}),
  \bibinfo{pages}{2048--2071}.
\newblock
\urldef\tempurl%
\url{https://doi.org/10.1111/j.1559-1816.2007.00250.x}
\showDOI{\tempurl}


\end{thebibliography}

\appendix

\section{Survey Questions}

{\itshape

\noindent Are you currently employed?

\vspace{.5em}

\noindent How many years of work experience do you have?

\vspace{.5em}

\noindent In the last year, how many jobs have you applied for?

\vspace{.5em}

\noindent (If one or more applications) In the last year, how many of your applications were unsuccessful?

\vspace{.5em}

\noindent (If one or more unsuccessful applications) Did any of the prospective employers explain why your application was unsuccessful?
\begin{itemize}
    \item (If yes) Please describe the most memorable explanation you received.
    \item (If no) How did you feel when you did not receive an explanation for an unsuccessful job application?
\end{itemize}

\vspace{.5em}
            
\noindent If you did not receive an explanation from an unsuccessful job application, would you be willing to apply for another position at the same employer?

\vspace{.5em}

\noindent Suppose an employer agreed to explain why your job application was unsuccessful. What information would you like the explanation to contain?

\vspace{.5em}

\noindent In a hypothetical scenario that you could purchase your ideal explanation from a past unsuccessful job application, how much would you be willing to pay for the explanation?

\vspace{.5em}

\noindent Explanations for job rejections may contain many different types of content. Please rate the importance of the following type of content.
\begin{itemize}
    \item Characteristics of the successful applicants (such as education level, years of experience, and main skills)
    \item How your profile would need to change to improve your chances.
    \item The main factors the employer used to distinguish between applicants.
    \item Gender of the successful applicant.
    \item Confirmation that the position was filled.
    \item Your ranking and the total number of applicants.
    \item Number of applicants who continued to the next round.
\end{itemize}

\vspace{.5em}

\noindent If your explanation could only contain one type of content, which type of content would you prefer (from the list above)?
}

\end{document}